# Super-rays grouping scheme and novel coding architecture for computational time reduction of graph-based Light Field coding

Nguyen Gia Bach [1], Chanh Minh Tran [1], Nguyen Duc Tho [1], Phan Xuan Tan [2,*], and Eiji Kamioka [1]

[1] Graduate School of Engineering and Science, Shibaura Institute of Technology, Tokyo 135-8548, Japan; Email: nb23505@shibaura-it.ac.jp (N.G.B.); nb20502@shibaura-it.ac.jp (C.M.T.); nb20501@shibaura-it.ac.jp (T.N.D.); kamioka@shibaura-it.ac.jp (E.K.)

[2] Department of Information and Communications Engineering, Shibaura Institute of Technology, Tokyo 135-8548, Japan;

*Correspondence: tanpx@shibaura-it.ac.jp.

*Abstract*—Graph-based Light Field coding using the concept of super-rays is powerful to exploit signal redundancy along irregular shapes and achieves good energy compaction, compared to rectangular block -based approaches. However, its main limitation lies in the high time complexity for eigen-decomposition of each super-ray local graph, a high number of which can be found in a Light Field when segmented into super-rays. This paper examines a grouping scheme for super-rays in order to reduce the number of eigen-decomposition times, and proposes a novel coding architecture to handle the signal residual data arising for each super-ray group, as a tradeoff to achieve lower computational time. Experimental results have shown to reduce a considerable amount of decoding time for Light Field scenes, despite having a slight increase in the coding bitrates when compared with the original non-grouping super-ray -based approach. The proposal also remains to have competitive performance in Rate Distortion in comparison to HEVC-based and JPEG Pleno -based methods.

*Keywords*—light field; compression; super-ray; over-segmentation; graph transform; eigen-decomposition

## I. Introduction

Light Field (Light Field) devices have recently been gaining popularity over traditional cameras for capturing light rays emitted by a 3D point from different orientations [1], hence providing a rich description of the 3D scene with a variety of potential applications in computer vision tasks like semantic segmentation [2], depth estimation and re-focusing [3], 3D reconstruction [4], video stabilization [5], and so on. This, however, comes with the cost of containing high dimensional data with redundancy in both spatial and angular dimensions, raising challenges in storage capacity and transmission [6].

Such redundancies or correlations can be visualized in a multi-view representation of a Light Field, illustrated in Fig. 1. Spatial correlation refers to the relation between the intensity values of nearby pixels in a single view. If the intensity values of two neighboring pixels are highly similar, they have a strong spatial correlation. Angular correlation refers to the relation between the light intensities of corresponding pixels from different views. If the intensities from different angles are highly similar, they have a strong angular correlation. The spatio-angular correlation has shown to be well exploited by signal transform-based methods [7], in which continuously similar signals with high redundancy are translated into energy coefficients with compaction in frequency domain. High energy compaction enables efficient compression performance.

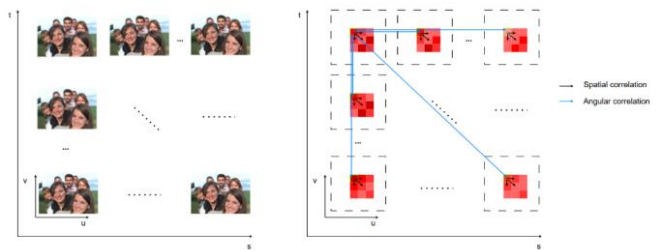

**Figure 1.** Multiview representation of a 4D Light Field scene L(u, v, s, t), with spatial domain (u, v), and angular domain (s, t). (Left side) Example of a Light Field scene Friends [8]. (Right side) Example of spatial correlation and angular correlation for a given pixel position in all views (selected in green)

This paper considers the use of graph-based transform to exploit the correlations in Light Field data. Graphs are useful to describe irregular image structures, adhering closely to texture boundaries. Thus, the image can be partitioned into variable-shape blocks containing mostly uniform pixel intensities. In contrast to the conventional rectangular blocks with variable-size used in standardized solutions (i.e: HEVC [9]), this is more likely to contain non-uniform intensities or different statistical properties. As a result, better energy compaction can be achieved for varying-sized blocks than rectangular blocks after transforming into frequency domain.

To carry out this operation, Graph Fourier transform (GFT) [10] and other variants [11–14] are natural tools used for adaptive transforms of irregular image structures, such as piecewise smooth images. A comprehensive survey on GFT can be found in [15]. GFT operation on each graph requires computing its basis functions (Laplacian eigenvectors) to decompose signals residing on a graph onto these basis functions. During this process, the Laplacian graph is diagonalized into eigenvalues and eigenvectors matrices, also known as eigen-decomposition. Nevertheless, when the dimension of the graph increases along with the dimension of signals, its computational complexity for basis functions rapidly becomes unfeasible, since the cost of computing eigenvectors is $O(n^3)$, with n as the number of vertices. This is the case when forming a full graph for high dimensional Light Field data, to efficiently exploit correlations in both spatial and angular dimensions. Such a graph connects all pixels within and across views, resulting in a significant number of vertices and edges. High dimensional graph with high computational complexity is one key downside for graph-based Light Field coding.

To make graph complexity become tractable, the concept of super-rays has been introduced to Light Field coding [16], so that local transforms can be applied to local graphs, instead of the original complete graph. The authors present super-rays as extension of super-pixels over-segmentation from 2D to 3D space, grouping similar nearby light rays coming from a surface in 3D space to different viewpoints, in a related manner that super-pixels group similar nearby pixels in 2D space.

This concept of super-rays forming local graphs enables further proposals on reducing complexity computing GFT basis functions like utilizing graph coarsening and partitioning for local graphs [17]. Graph coarsening lowers the graph dimension by reducing the number of vertices below a given threshold while retaining its basic graph properties, under an acceptable complexity. Additionally, after graph coarsening, the number of vertices is the same for many super-rays, leading to the same graph dimension, and some of the coarsened super-rays are also likely to share similar graph properties.

This might be related to the fact that, while super-rays group together similar light rays adjacent to each other, there might also exist similar light rays far from each other that have not been grouped. This raises a question about whether one can also group together super-rays, either close or far from each other, which might share similar signals or possibly spectral properties (i.e: their GFT basis functions). When two arbitrary super-rays have graph supports with similar basis functions, it is possible to do an Inverse Graph Fourier Transform (I-GFT) using one's basis functions for the other, then obtain a slightly different predicted signal for the other super-ray. The residual between the original and predicted signals can be small and contain redundancy, hence it is suitable for transmission.

Therefore, this paper introduces two proposals, 1) a super-rays grouping scheme for coarsened super-rays to decrease the number of times computing GFT basis functions, and 2) a coding architecture suitable for the grouping scheme and the transmission of residual signals. Regarding the grouping scheme, the super-rays are grouped together based on their similar coefficients after the GFT operation, then the main super-ray is selected in such a way that leads to small residual signals for the transmission. Regarding the coding architecture, at the decoder side, only one eigen-decomposition of the main super-ray is required to be executed in one group, instead of computing eigenvectors for every super-ray. Hence, this would save the time necessary to compute basis functions for every other super-rays in a group. Additionally, this paper also investigates the use of GPU and parallel computing for faster Laplacian diagonalization. The experimental results demonstrate that by using the proposed super-rays grouping scheme, computational time at the decoder side is considerably reduced for all Light Field with simple and complex scenes, and the use of GPU and parallelism has made the practical use of graph-based Light Field coding more feasible.

The rest of the paper is organized as follows: Section 2 introduces categories of Light Field compression and recent studies on graph-based Light Field compression. A detailed description of the proposed super-ray grouping scheme and coding architecture is given in Section 3. Section 4 provides experimental results and evaluates grouping performance on the time saving and Rate Distortion performance. Section 5 discusses overall results, and some limitations. The conclusion and future work are given in Section 6.

## II. LITERATURE REVIEW

This section first introduces current progress on Light Field compression, and where the graph-based approach fits into the compression families, then goes further into recent studies on reducing computational time for graph-based Light Field coding.

### A. Recent trends on Light Field Compression

The existing solutions can be generally classified into the following categories: raw lenslet (2D image) -based compression and multiple views (array of 2D sub-apature images extracted from the lenslet format) -based compression. An in-depth survey can be found in [7].

Most studies in the first category rely on adding novel prediction modes for existing codecs to exploit spatio-angular redundancy. For instance, authors in [18–23] expand HEVC Intra prediction modes with block-based self-similarity techniques. Extensions to other standards have also been considered like using JPEG-2000 [24] to encode residual data after sparse prediction for micro-images based on the depth map. In addition to these standards, [25] implements graph lifting transform technique for irregularly spaced color components in raw lenslet images without demosaicing. However, lenslet-based Light Field representation presents inconsistent pixel correlations, with a regular pattern of spikes found in the autocorrelation function, which is not as smooth as traditional 2D images [7]. Non-smooth correlations in the spatial domain results in lower energy compaction in the frequency domain, and thus it is the main challenge for the methods utilizing existing coding standards to compress 2D lenslet image.

Methods in the second category aim at compressing a set of views instead of a single 2D image, processing signals in all dimensions, including spatial, inter-view, and pseudo temporal domains. The diversity of ways these views are stacked together motivates various Light Field compression approaches. Pseudo-video-sequence (PVS) based methods scan 2D array of viewpoints to form a 1D array of views. This exploits (pseudo) temporal relation between views, similar to conventional 2D coder exploiting inter-frame correlation, i.e: using HEVC in [26], [27], or JEM coder [28]. Multiview based methods stack a 2D array of views into a 1D array of multiple PVSs as a 3D multiview format, then any conventional 3D video coders like MVC or MVC-HEVC [29, 30] can be used to process the views. Another approach follows a hierarchical order by first coding a sparse set of views in the base layer with HEVC, then predicting the views in the enhancement layer [31]. Nevertheless, these methods still rely on existing coding standards, exploiting pseudo correlations limited by the scanning order of viewpoints, and thus the intrinsic correlations of a Light Field scene might not yet be fully exploited.

Among the methods in the second category, the recent approach for Light Field compression with the assistance of geometry information does not depend heavily on how the viewpoints are stacked or trying to convert their representation into a 2D/3D video. Hence, geometry-assisted based Light Field representation relies less on conventional coders. Instead, existing studies on this approach focus on the selection of key views along with geometry estimation problems [32, 33], as illustrated in Fig. 2. The representation is accompanied by view synthesis based Light Field compression. JPEG Pleno, a new standard made by ISO/IEC JTC 1/SC 29/WG 1 JPEG Committee with specialization in novel image modalities like Light Field, point cloud or holograms, adopts the geometric-assisted method in the 4D Prediction mode (4DPM) [34]. This mode partitions the views into a set of reference views and a set of intermediate views. The decoder side utilizes a hierarchical depth-based prediction technique to obtain geometry information like depth maps, then warps missing textures for intermediate views from reference views based on depth information. In addition, JPEG Pleno also implements another mode called 4D Transform (4DTM) [34]. 4DTM partitions Light Field into 4D blocks of variable sizes, covering two spatial and two angular dimensions, then transformed each block using 4D DCT. JPEG Pleno 4DTM was selected for comparison in this work due to the unavailability of 4DPM in the open source of JPEG Pleno Reference Software at the time of writing this paper. It should be noted that rectangular pixel blocks are still used in these methods, which might contain non-uniform intensities and less redundancy to be exploited.

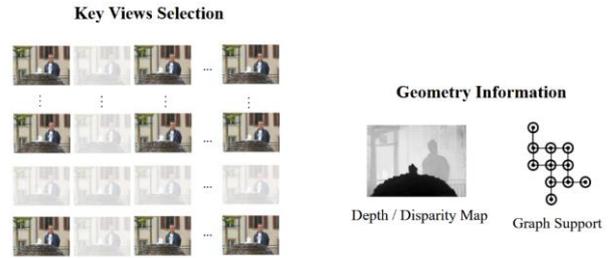

**Figure 2.** Geometry-assisted Light Field coding. (Left side) Key views are selected to estimate geometry information (Right side), such as depth/disparity, and graph model.

B. *Graph-based Light Field Coding*

Graph-based Light Field coding also belongs to the second category of Light Field compression, coding a dense array of 2D sub-aperture images extracted from lenslet Light Field, with the assistance of geometry information. Graph vertices represent pixels in a sub-aperture image along with color intensities as graph signals, and graph edges indicate correlations between pixels intra-view or inter-view. While the graph edges can be used to form Laplacian matrices and compute graph basis functions to support GFT, graph signals are transformed into coefficients in frequency domain using GFT to exploit spatio-angular redundancy.

The proposals from [35] and [36] first segment the top-left view of Light Field into super-pixels using the SLIC algorithm [37], and compute its disparity map, as illustrated in Fig. 3. Subsequently, super-pixel labels are projected from the top-left view to the remaining views based on the median disparity per super-pixel using the projection scheme depicted in Fig. 4a. A set of corresponding super-pixels in all the views form a super-ray, which also represents a local graph, containing spatial graphs with vertices as pixels in every super-pixel per view, and angular graphs connecting corresponding pixels between super-pixels in different views. Fig. 5 provides an example of a local graph. However, diagonalizing a complete graph of both spatial and angular graphs is expensive, and thus the authors carried out GFT in a separable manner for each super-ray graph, spatial transform to exploit spatial correlation within a super-pixel, and angular transform to exploit inter-view correlation of corresponding super-pixels. [17] proposed another solution to reduce the eigen-decomposition time of high dimensional graphs by trying to reduce super-ray size in a rate distortion sense, using graph coarsening and partitioning. Their results have outperformed state-of-the-art coders on ideal Light Field (real Light Field with small parallax), but performed worse than coders like HEVC Lozenge [26] and JPEG Pleno [38] at high bitrates due to vignetting effect in real Light Field with high parallax, and synthetic Light Field with large disparity. A previous study of this paper's authors [39] attempted to solve these issues using center-view projection and multiple-view projection schemes for the two types of Light Field scenes, as illustrated in Fig. 4b and Fig. 4c. Rate-distortion performance using the two projection schemes has been shown to increase significantly compared to the original top-left view projection scheme. Nevertheless, computational time remains to be high, which is a challenge for graph-based Light Field coding, and thus this paper proposes a novel super-rays grouping scheme to deal with this problem.

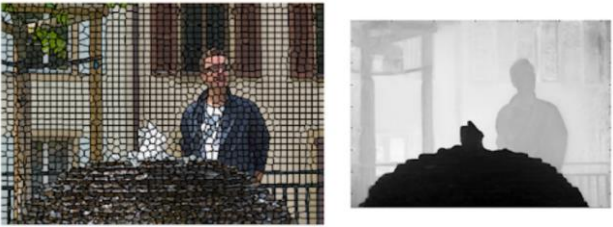

**Figure 3.** (Left side) Super-pixels over-segmentation of a view using SLIC algorithm (Right side) Estimated disparity map using [40]

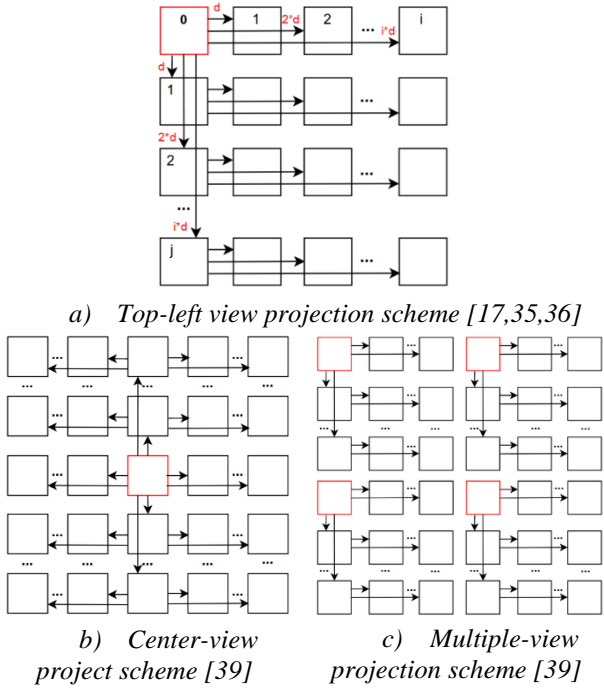

*a) Top-left view projection scheme [17,35,36]*

*b) Center-view project scheme [39]*  *c) Multiple-view projection scheme [39]*

**Figure 4.** Super-rays projection schemes

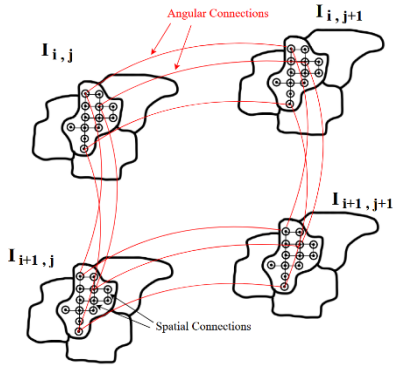

**Figure 5.** Local graph construction for a super-ray, connecting pixels inside every super-pixel, and corresponding pixels between super-pixels

## III. PROPOSALS

Since a high number of super-rays may incorporate a Light Field, leading to a high number of eigen-decompositions, in this section, a super-rays grouping scheme is proposed to reduce the number of times for this operation. The grouped super-rays should share similar coefficients after GFT, in such a way that the predicted signals are close to the original after I-GFT using the same basis functions for all local graphs in one group. Subsequently, a coding scheme for both the encoder and decoder is proposed to handle the residual signals after prediction, and other necessary data to be transmitted such as geometry information, segmentation labels, and transform coefficients.

### A. Problem Statement

Denote a graph as $G = \{V, E\}$, corresponds to its signal f. Its adjacency matrix is denoted as A where $A_{i,j} = 1$ if there is an edge between vertex $v_i$ and $v_j$, and $A_{i,j} = 0$ otherwise. Its degree matrix D is diagonal, in which $d_{i,i} = \sum_j A_{i,j}$. Its Laplacian matrix $L = D - A$ is symmetric positive semi-definite, and thus L can be diagonalized (eigen-decomposed) as in Eq. (1)

$$L = U\Lambda U^T \qquad (1)$$

, in which $\Lambda$ is diagonal matrix made of real positive eigenvalues $\lambda_k$ representing graph frequencies, while U matrix consists of orthogonal eigenvector columns (basis functions) interpreted as graph frequency components corresponding to each eigenvalue. Eigenvectors matrix U of a graph G is used for Graph Fourier Transform (GFT) operation to decompose signal f lying on G into its frequency coefficients by the formula $\hat{f} = U^T f$, in a similar manner to Discrete Cosine Transform (DCT) but with the ability to apply on irregular regions of highly smooth signals.

The main issue here is high time complexity $O(n^3)$ diagonalizing a $n \times n$ Laplacian matrix to obtain the basis functions U (eigenvectors). That is the computational cost for only 1 local graph (super-ray), however, Light Field can be decomposed into a high number of super-rays, even more with graph partitioning to obtain higher quality. For instance, a Light Field scene *Greek* [41], composing 9×9 views, can be initially segmented into 700 super-rays (700 super-pixels per view), as illustrated in Fig. 6a, but rising up to 4390 super-rays when enabling graph coarsening and partitioning, as shown in Fig. 6b. Each super-ray's eigen-decomposition may take up to 15s (seconds) under a high-end CPU, for a local graph of 5000 vertices to obtain an acceptable quality [17]. Thus, it may lead up to about 15∗4390 = 65850 seconds, approximately 18.3h (hours), to process the whole *Greek* scene of 4390 super-rays if the eigen-decompostions are carried out sequentially. If there is a solution to avoid the number of times necessary for eigen-decompositions, the overall computing time may be reduced accordingly. One potential method is to group similar super-rays together, then carry out eigen-decomposition for only one main super-ray per group. As explained in the next section, the "similar super-rays" are defined in this paper as those having similar coefficients after the GFT operation, and the main super-ray per group should be selected in such a way as to obtain minimal signal residual of each group for easier transmission.

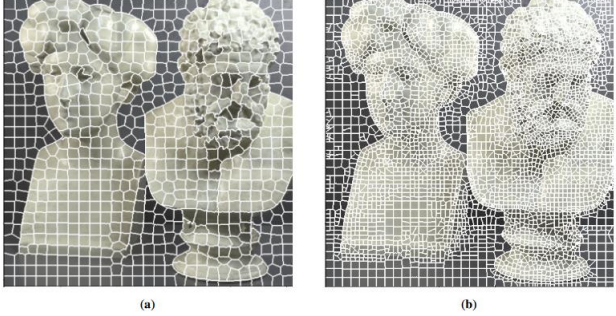

**Figure 6.** SLIC segmentation of one view in dataset Greek [41] into super-pixels. (a) Original segmentation map (b) Optimized segmentation map after graph coarsening and partitioning

### B. Super-rays grouping scheme

Assume a group of super-rays is already formed, denoted as $\{SR_{i,j}\}$, in which $i$ is index of a group, and $j$ is the index of super-ray in that group. Each super-ray represents a local graph, with Laplacian matrices denoted as $\{l_{i,j}\}$, and the pixel signals lying on the graph $\{f_{i,j}\}$. $\{l_{i,j}\}$ can be eigen-decomposed into eigenvalues $\{e_{i,j}\}$ and eigenvectors $\{u_{i,j}\}$ using Eq. (1).

For each group of super-rays, assume that an arbitrary super-ray is selected as the main one, denoted as $SR_i$, along with its Laplacian matrix $l_i$, and its signals $f_i$. $l_i$ can be diagonalized using Eq. (1) to obtain eigenvalues $e_i$ and eigenvectors $u_i$.

$$u_{i,j}^T f_{i,j} = \hat{f}_{i,j} \qquad (2)$$
$$f_{i,j} = \hat{f}_{i,j} \qquad (3)$$
$$u_i \hat{f}_{i,j} = f'_{i,j} \qquad (4)$$

From here, the goal is to achieve $f'_{i,j}$ as similar to $f_{i,j}$ as possible, so the residual $\Delta \hat{f}_{i,j} = f_{i,j} - f'_{i,j}$ would be small to be sent to the decoder side, along with only the main super-ray's eigenvectors $u_i$ and coefficients $\hat{f}_{i,j}$. This allows the reconstruction of the original signal $f_{i,j} = \Delta \hat{f}_{i,j} + f'_{i,j}$ at the decoder side, without having to calculate $u_{i,j}$. This problem can be interpreted as the following equation:

$$\begin{aligned}
&\min\left(f_{i,j} - f'_{i,j}\right) \\
\Leftrightarrow\ &\min(f_{i,j} - f_i) + \min\left(f_i - f'_{i,j}\right) \\
\Leftrightarrow\ &\min(f_{i,j} - f_i) + \min(u_i * \hat{f}_i - u_i * \hat{f}_{i,j}) \\
\Leftrightarrow\ &\min(f_{i,j} - f_i) + \min(\hat{f}_i - \hat{f}_{i,j}) \qquad (5)
\end{aligned}$$

This means for every group of super-rays, a main super-ray should be chosen in attempt to satisfy these equations:

$$\begin{cases} \min(f_{i,j} - f_i) & (a) \\ \min(\hat{f}_i - \hat{f}_{i,j}) & (b) \end{cases}$$

Condition (a) minimizes the difference between coefficients after GFT of main super-ray and every other super-ray in a group, meanwhile condition (b) minimizes the difference between their original graph signals.

Therefore, in the beginning, given a set of super-rays not having been grouped, a super-rays grouping scheme is proposed as follows:

- Step 1: group super-rays with similar coefficients into one group based on an optimal MSE threshold, following condition (a)
- Step 2: following condition (b), for each group of super-rays, select a main super-ray candidate having graph signal values $f_i$ as the median signal of all super-rays $\{f_{i,j}\}$ in the group. The median value $f_i = median(\{f_{i,j}\})$ allows the set of residual signals $\{f_{i,j} - f_i\}$ to avoid having high values, while more likely to contain similar values (high redundancy), leading to high redundancy in the final residuals $\Delta \hat{f}_{i,j} = f_{i,j} - f'_{i,j}$, which is efficient for transmission.

Going further into step 1, the grouping scheme can be broken down into smaller steps, as described in Fig. 7. Intuitively, one can consider each super-ray as a graph node, and a set of arbitrary super-rays is represented as a graph with no edges initially, as in Fig. 7a. Denote total number of coarsened super-rays as m.

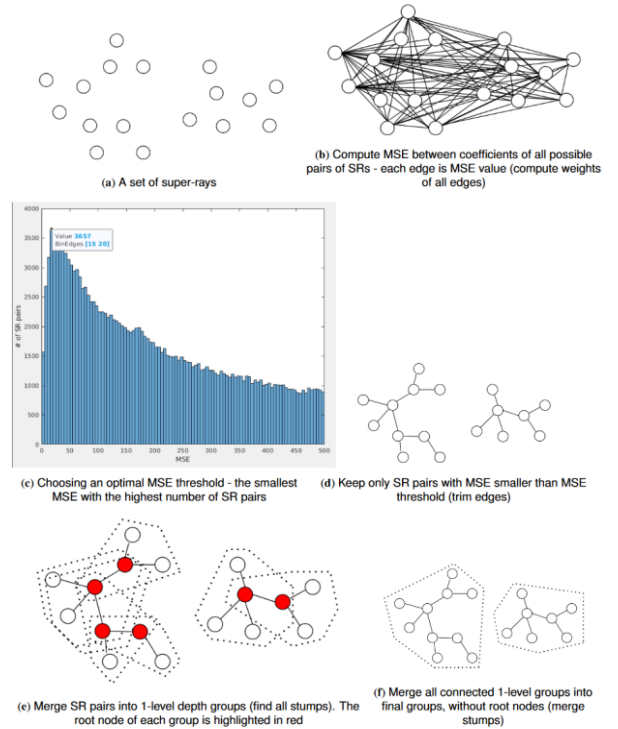

**Figure 7.** Visualization of super-rays grouping scheme

Then, an edge for every pair of super-rays will be drawn, with the weight calculated as the MSE score between their coefficients, as shown in Fig. 7b. The total number of pairs is m combinations of 2, $C(m, 2)$. Once MSE values of all pairs are computed, an optimal MSE threshold will be chosen based on a histogram, summarizing distribution of the number of super-ray pairs over MSE bins. In this histogram, the number of pairs is expected to be as high as possible, meaning that more super-ray pairs can satisfy condition (a) with respect to a MSE. Meanwhile, the MSE bins should be small, since high values of MSE will lead to high residuals of signals $f_{i,j} - f'_{i,j}$, meaning less redundancy data to be exploited, and its compression becomes less efficient. Therefore, the optimal MSE threshold is chosen in a way that it is the smallest MSE bin with the highest number of super-ray pairs. An example is shown in Fig. 7c. Once obtained the optimal MSE threshold, only the edges with MSE weight smaller than the threshold are kept, whereas the remaining edges are removed, as illustrated in Fig. 7d.

After super-ray pairs with similar coefficients are selected, the grouping process starts by merging all pairs into 1-level depth groups, which means all remaining super-rays in this group are connected to only the main super-ray, as shown in Fig. 7e. This can be done by iterating the list of similar super-rays, for each one, grouping all other ones similar to that super-ray, having MSE score smaller than or equal to the optimal threshold.

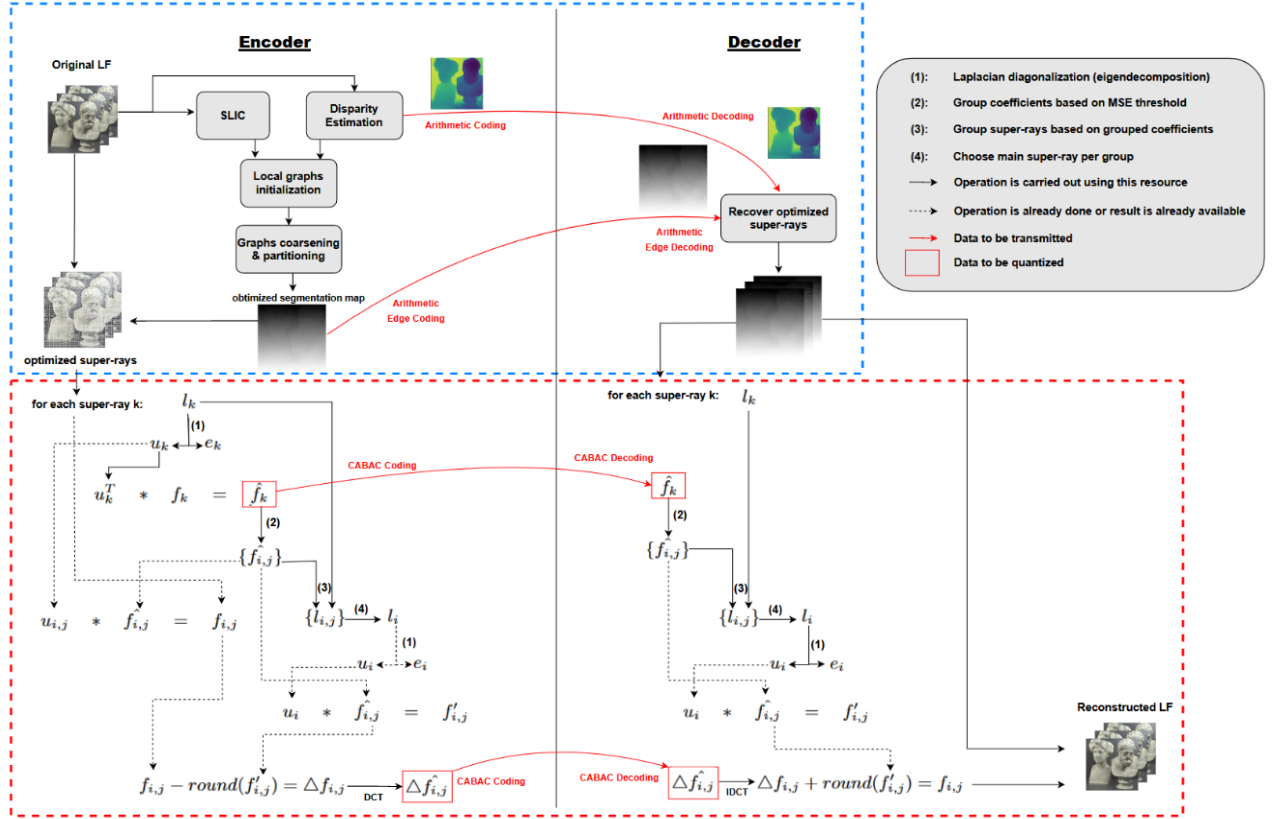

**Figure 8.** Proposed Light Field coding architecture. Steps (2) and (4) correspond to super-rays grouping scheme (algorithm 1) and the selection of main super-ray respectively in Section 3.2. The part highlighted in blue corresponds to the proposed architecture in [17]. The part highlighted in red corresponds to this paper's proposals.

Finally, all 1-level depth groups having intersections are grouped into final d-level depth groups, described in Fig. 7f. The algorithm for step 1 is detailed in Algorithm 1. The time complexity for this algorithm is $O(C(m, 2))$ or $O(m!/(2! * (m - 2)!))$, with m as the number of coarsened super-rays, which is usually much smaller than the number of vertices in a local graph, affecting the time complexity of graph Laplacian diagonalization $O(n^3)$.

**Algorithm 1** Coarsened super-rays grouping algorithm

**Input:** A set of coarsened super-rays coefficients: *coarsened_coeff*
**Output:** Optimized groups of similar coarsened super-rays
**Initialize:**
    *mse_weights* : hashmap
    *sub_groups* ← ∅
    *final_groups* ← ∅
**for** i ← 1 to *coarsened_coeff_length* **do**
    **for** j ← 1 to *coarsened_coeff_length* **do**
        **if** j ≠ i AND *mse_weights{i, j}* is 0 **then**
            *mse_weights{i, j}* ← mse(*coarsened_coeff*[i], *coarsened_coeff*[j])
        **end if**

```
        end for
    end for

    mse_threshold ← first_max (hist(mse_weights))

    for i ← 1 to coarsened_coeff_length do
        sub_groups[i] ← find(mse_weights{i}
                              ≤ mse_threshold)
    end for

    for each sub_group in sub_groups do
        for each final_group in final_groups do
            if intersect(sub_group, final_group) ≠ ∅ then
                final_groups[i] ← sub_group ∪
                    final_group
            end if
        end for
        if final_groups is empty then
            final_groups ← [sub_group]
        end if
    end for
```

### C. Coding architecture

Fig. 8 provides a detailed description of the proposed coding architecture. At the encoder side, the top left view of the original Light Field is first segmented into super-pixels using the SLIC algorithm [37]. Then, the disparity map is estimated from the Light Field with the method used in the original paper [40]. These segmentation and disparity maps are used to initialize local graphs by projecting super-pixels from the top-left view to the remaining views, then super-rays are obtained, containing sets of corresponding super-pixels in all views. Each super-ray represents a local graph, and these graphs are further reduced by graph coarsening and partitioning in a rate-distortion sense, as proposed in [17]. As a result, optimized segmentation maps (super-rays) are obtained, corresponding to their optimized local graphs. The segmentation labels are then coded along with disparity values for each super-ray, ready to be sent to the decoder side. This part is similar to the previous architecture proposed in [17].

Next, this section proposes novel modifications to the existing architecture. For each super-ray $k$, each corresponding local graph $l_k$ will be diagonalized to derive eigenvectors $u_k$, which is then used for GFT with its graph signals $f_k$ to compute graph coefficients $\hat{f}_k$. These coefficients will be used in Algorithm 1 to obtain groups of similar super-rays $l_{i,j}$. For each group, a main super-ray $l_i$ is chosen having median signal. Importantly, its pre-computed eigenvectors $u_i$ are then used to do I-GFT with coefficients $\hat{f}_{i,j}$ of each remaining super-ray in that group, resulting in a slightly different predicted signal $f'_{i,j}$ for each super-ray. The 1-D residuals $\Delta\hat{f}_{i,j}$ between 1-D original $f_{i,j}$ and rounded predicted signals $f'_{i,j}$ are first transformed into coefficients using 1-D DCT. Both GFT and DCT coefficients are then quantized, and entropy coded using a public version of CABAC (Context-based adaptive binary arithmetic coding) [42].

At the decoder side, it receives the decoded optimized segmentation labels of the first view along with disparity information per super-pixel, then carries out super-pixels projection to obtain segmentation maps for remaining views, resulting in optimized super-rays. Subsequently, the decoded GFT coefficients $\hat{f}_k$ are used to group similar super-rays $l_{i,j}$ using Algorithm 1 in a similar manner as encoder. For each group, the eigenvectors $u_i$ of the main super-ray $l_i$ are used again for I-GFT with coefficients $\hat{f}_{i,j}$ of each remaining super-ray, obtaining predicted signal $f'_{i,j}$. Adding the rounded $f'_{i,j}$ with corresponding residuals $\Delta\hat{f}_{i,j}$ received from the encoder, the reconstructed signal $f_{i,j}$ for each super-ray is then obtained. Afterwards, the Light Field is retrieved using the segmentation maps of local graphs and their corresponding reconstructed signals.

Importantly, it should be noted that, at the encoder side, only GFT coefficients of non-grouped super-rays and main super-rays of each group are coded and transmitted, instead of all super-rays. At the decoder side, eigen-decomposition operation is only required for non-grouped super-rays and main super-rays, potentially saving a significant amount of time computing basis functions for the grouped super-rays, depending on the grouping ratio of how many super-rays are grouped over the total number of coarsened super-rays.

## IV. EXPERIMENT RESULT AND DISCUSSION

This section examines the results of applying the proposed super-rays grouping scheme, and coding architecture on four challenging light fields. A set of experiments are designed on the Encoder and Decoder machines to evaluate how much time saving can be achieved when only one eigen-decomposition is needed per group of super-rays, using GPU to accelerate diagonalization of high dimensional Laplacian matrices. Additionally, rate distortion performance of the proposed super-rays grouping and original graph-based Light Field coding are also compared against state-of-the-art coders, to illustrate the tradeoff between slightly increasing bitrate and gaining a considerable amount of time saving at the decoder side.

### A. Experiment Setup

The datasets used for evaluation were selected following the Light Field Common Test Conditions Document [43] to obtain a variety of scenes that the proposed super-rays grouping scheme and compression algorithms would challenge, in terms of capturing technology (plenoptic camera or computer generating), scene geometry and texture, resolutions, number of viewpoints, bit depth. The chosen real Light Fields containing natural scenes Fountain_Vincent_2 and Danger_de_Mort were obtained from the EPFL dataset [8], which was acquired using a Lytro Illum camera. The scenes were preprocessed using Light Field Toolbox [44] on Matlab to obtain 15x15 views of 625x434 resolution at 10-bit depth in each view, but only 13x13 views were extracted and applied with gamma correction to avoid strong vignetting effect. The synthetic Light Fields selected in this paper were Greek and Sideboard from HCI 4D Light Field dataset [41], containing photorealistic scenes of 9x9 views at 512x512 resolution and 8-bit depth, generated with Blender software.

The Encoder and Decoder were run on Python 3 under Ubuntu 20.04 with 64GB RAM, 11th Gen Intel(R) Core(TM) i9, 3.50GHz clock speed, NVIDIA GeForce RTX 3080 with 10GB video memory. The experiments utilized Python's Ray library [45] for the parallel Laplacian diagonalization of super-ray local graphs on both CPU and GPU scenarios. The GPU scenario used Python's Scikit-cuda library [46] to manipulate the data on CUDA device/runtime and eigen-decompose the matrices with cuSOLVER algorithm, and a GPU accelerated library for linear system solutions and decompositions for both sparse and dense matrices [47].

### B. Analysis of super-rays grouping performance

Table 1 summarizes grouping results using Algorithm 1 on synthetic Light Fields *Greek* & *Sideboard* with 9x9 views, and plenoptic Light Fields *Fountain_Vincent*2 & *Danger* with 13x13 views, at high quality (high bitrates) coding. The grouping is analyzed in terms of the number of super-rays (*# SRs*), the number of coarsened super-rays (*# coarsened SRs*) corresponding to the number of nodes in Fig. 7a, the number of super-ray pairs (*# SR pairs*) corresponding to the number of edges in Fig. 7b, the optimal MSE threshold (*mse_threshold*) corresponding to Fig. 7c, the number of super-ray pairs after being reduced with respect to mse_threshold (*# SR pairs w.r.t mse_threshold*) corresponding to Fig. 7d, the number of 1-level depth groups (*# 1-level groups*) corresponding to Fig. 7e, the number of final groups (*# final groups*) corresponding to Fig. 7f, the number of coarsened super-rays having been grouped (*# grouped coarsened SRs*), the grouping ratio for coarsened super-rays (*coarsened grouping ratio*), and the grouping ratio for both coarsened and partitioned super-rays (*overall grouping ratio*). It should be noted that the number of super-rays is smaller in 9x9 views Light Field than 13x13. Light Fields with larger views are partitioned into more super-rays because a higher number of views lead to more super-pixels in one super-ray, while its total number of vertices is fixed for coarsened super-rays, and thus super-pixel boundaries are smaller in each view, leading to more super-rays to being formed. Additionally, for a high target quality PSNR coding, signal approximation on the coarsened graph may return a too coarse approximation of the original signal, hence graph partitioning is applied instead of coarsening, splitting each super-ray into smaller ones [17]. This can be shown in the low number of coarsened super-rays in the table.

| Datasets | Greek | Sideboard | Fountain_Vincent2 | Danger |
|---|---|---|---|---|
| # SRs | 4390 | 5993 | 17616 | 12253 |
| # coarsened SRs | 1252 | 853 | 1659 | 2723 |
| # SRs pairs | 783126 | 363378 | 1375311 | 3706003 |
| mse_threshold | 20 | 10 | 45 | 50 |
| # SRs pairs, w.r.t mse_threshold | 3657 | 730 | 8627 | 2173 |
| # 1-level groups | 382 | 201 | 548 | 329 |
| # final groups | 21 | 13 | 20 | 6 |
| # grouped coarsened SRs | 1026 | 418 | 1013 | 763 |
| coarsened grouping ratio | 0.82 | 0.49 | 0.61 | 0.28 |
| Overall grouping ratio | 0.23 | 0.07 | 0.06 | 0.06 |

**Table 1.** Results on super-rays grouping scheme using Algorithm 1

Coarsened grouping ratio, as defined in Eq. 6, reflects the true grouping performance without considering partitioned super-rays, which is considered by the overall grouping ratio, defined in Eq. 7. Having a smaller number of super-rays may lead to a higher overall grouping ratio due to smaller number of partitioned super-rays, while coarsened grouping ratio is not affected

$$coarsened\_grouping\_ratio = \frac{\# \; grouped\_coarsened\_SRs}{\# \; coarsened\_SRs} \quad (6)$$

$$overall\_grouping\_ratio = \frac{\# \; grouped\_coarsened\_SRs}{\# \; SRs} \quad (7)$$

Regarding step 1 in the grouping scheme, Fig. 9 displays the histogram of coefficients MSE and its corresponding number of super-rays pairs, along with the chosen threshold value. mse_threshold considerably reduces the number of uncorrelated super-rays pairs, while keeping a high number of pairs with similar coefficients.

Since Greek and Fountain_Vincent_2 Light Field scenes are less complex, more coarsened super-rays could be grouped, and their coarsened grouping ratios were relatively higher than Sideboard's and Danger_de_Mort's. However, all four datasets' overall grouping ratio remained low, since most super-rays were partitioned in high PSNR coding, and thus the grouping scheme wasn't applied to these super-rays.

In summary, coarsened super-rays grouping reduces the number of eigen-decomposition times from 1026, 418,

1013, 763 to only 21, 13, 20, 6 on datasets Greek, Sideboard, Fountain_Vincent_2, Danger_de_Mort respectively. If these operations are carried out sequentially on coarsened super-rays of 5000 vertices, assuming each takes around 15s using the current experiment setup on CPU environment, this grouping scheme can save about 15075s (approx. 4.2h), 6075s (approx. 1.7h), 14895s (approx. 4.1h), 11355s (approx. 3.2h) on each dataset correspondingly.

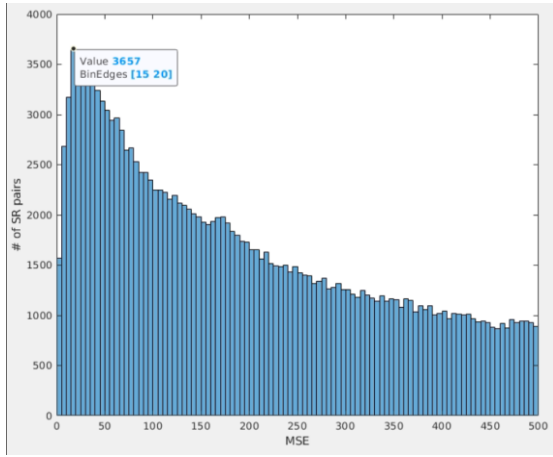

*a)   Greek*

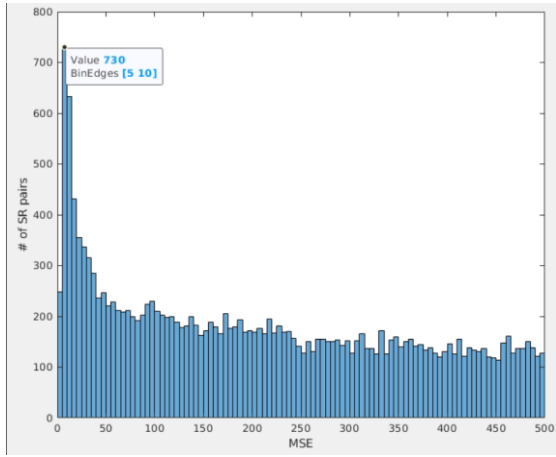

*b)   Sideboard*

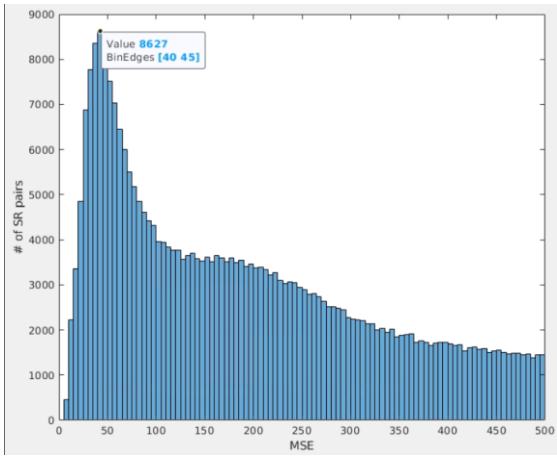

*c)   Fountain_vincent2*

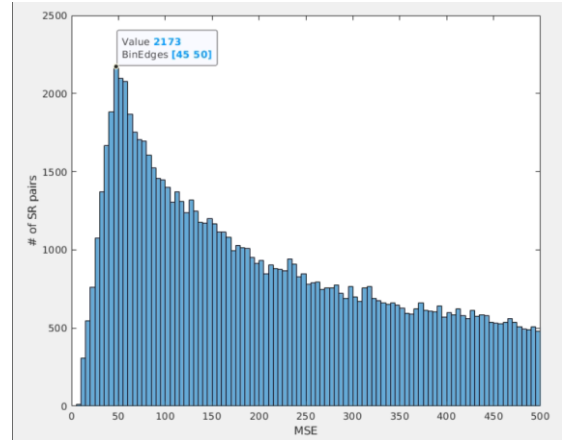

*d)   Danger_de_mort*

**Figure 9.** Histogram of coefficients MSE and corresponding number of super-ray pairs on four Light Field scenes. The vertical axis represents the MSE values between pairs. The selected MSE threshold for each scene is highlighted in each histogram.

Fig. 10 displays four example final groups of coarsened super-rays in the four datasets. Each row represents four final super-ray groups in a dataset and each column represents one final group in each dataset. For each final super-ray group, the grouped super-pixels (grouped super-rays in one view) are highlighted in red. The proposed scheme has the ability to group together super-rays in regions with highly similar color signals. Even though the shapes of these grouped super-rays can be inconsistent, their underlying coarsened graphs might share similar basis functions, and thus they can be grouped using Algorithm 1 based on their similar coefficients. Particularly, there are many large groups on Greek and Fountain_Vincent_2 datasets due to having more uniform regions, in which the residing super-rays may share similar signals and underlying graph spectral properties. Meanwhile, complex Light Field scenes like Sideboard and Danger_de_Mort contain fewer uniform textures, limiting the grouping ability. It should also be noted that the grouped super-rays adhere closely to the uniform region boundaries, which might be considered as an extension to the concept of super-rays / super-pixels grouping similar adjacent light rays / pixels in irregular regions.

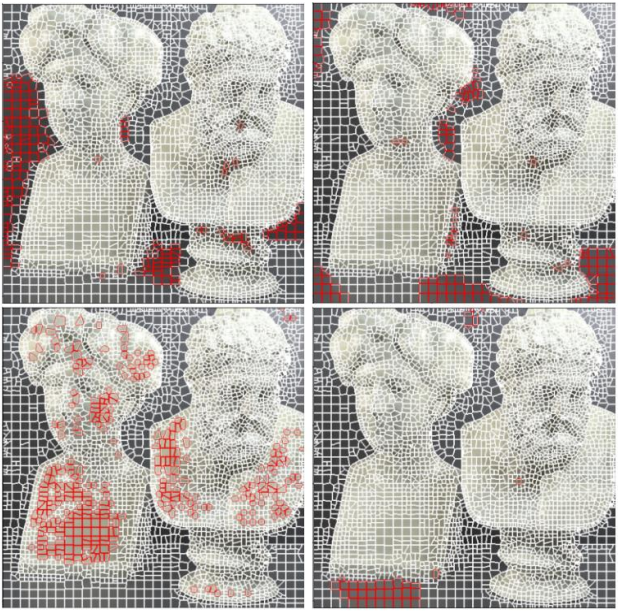

(a) Greek

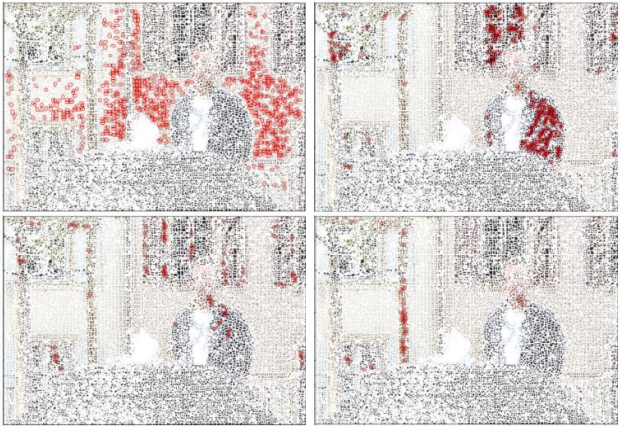

(b) Fountain_Vincent2

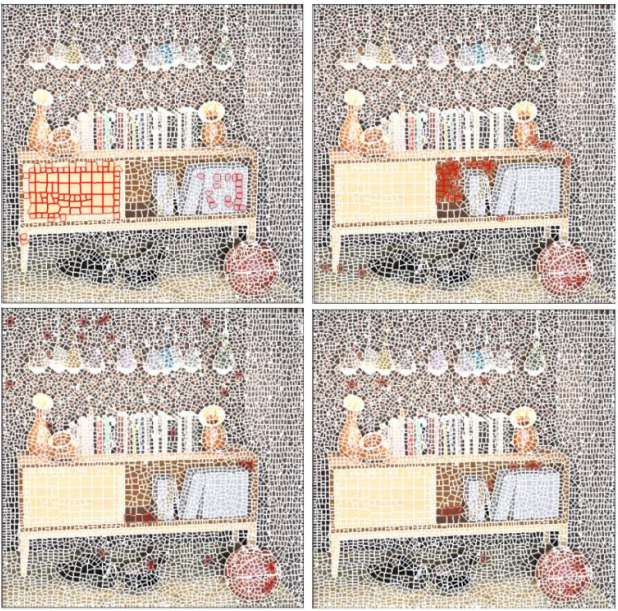

(c) Sideboard

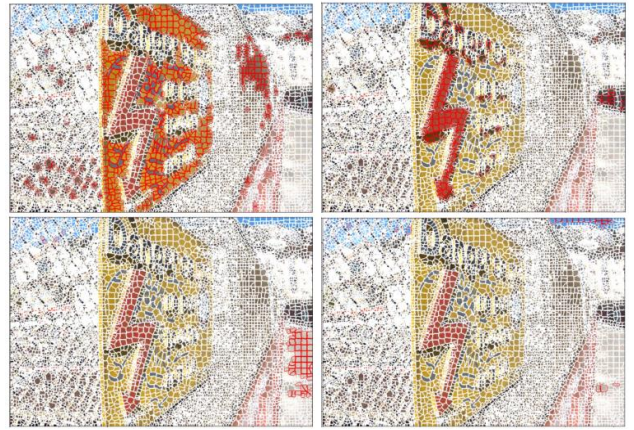

(d) Danger_de_Mort

**Figure 10.** Example of four final groups of coarsened super-rays in four datasets: (a) Greek, (b) Fountain_Vincent_2, (c) Sideboard, (d) Danger_de_Mort. Each subfigure in a dataset represents one final group in that light field.

C. *Analysis of Coding performance*

2) *Encoding and Decoding time*

Tables 2, 3, 4, and 5 demonstrate the computational time results on the four Light Field scenes when running the Encoder and Decoder with the use of Python's Ray library for parallel processing of multiple super-rays under the given experimental setup environment. Both Encoder and Decoder were evaluated under two scenarios, with or without GPU support for eigen-decompostion operation of local graphs. In each scenario, the running times of the two machines were compared between the original approach [17] and the proposed grouping approach. It should be noted that the encoding times for both approaches are the same for each scenario, because in the architecture proposed in Fig. 8, eigen-decomposition is still required for every super-ray at the encoder side to obtain their basis functions, which are then used for GFT operation to compute coefficients and group super-rays. The time saving from the tables can be interpreted for each dataset as follows: grouping proposal saves around 11278s (3.1h) for no-GPU scenario or 2355s (0.7h) for with-GPU scenario on Greek; 4710s (1.3h) or 1631s (0.5h) respectively on Sideboard; 8789s (2.4h) or 3150s (0.9h) respectively on Fountain_Vincent_2; and 7950s (2.2h) or 2661s (0.7h) respectively on Danger_de_Mort. The time saving corresponds well to the overall grouping ratio results in Table 1 and the size of Light Field (number of views). Greek had the highest grouping ratio due to its moderately smooth Light Field scene, hence saving a good amount of time. Although Sideboard had a slightly better grouping ratio than Fountain_Vincent_2 and Danger_de_Mort, it required less time processing due to having a smaller number of views, and thus time saving in the other two real Light Fields are more significant. Additionally, GPU support has shown to significantly reduce processing time for both Encoder and Decoder, saving 17725s (4.9h), 9022s (2.5h), 30397s (8.4h), 27951s (7.8h) on Greek, Sideboard, Fountain_Vincent_2, Danger_de_Mort respectively at the Encoder side.

|  | **no-gpu** |  | **with-gpu** |  |
|---|---|---|---|---|
|  | original | grouping | original | grouping |
| **Encoder** | 35690 (s) |  | 17965 (s) |  |
| **Decoder** | 31029 (s) | 19751 (s) | 8811 (s) | 6456 (s) |

**Table 2.** Computational time comparing grouping scheme and no-grouping scheme in both encoder and decoder on dataset Greek

|  | **no-gpu** |  | **with-gpu** |  |
|---|---|---|---|---|
|  | original | grouping | original | grouping |
| **Encoder** | 97250 (s) |  | 66853 (s) |  |
| **Decoder** | 60677 (s) | 51888 (s) | 21641 (s) | 18491 (s) |

**Table 3.** Computational time comparing grouping scheme and no-grouping scheme in both encoder and decoder on dataset Fountain_Vincent_2

|  | **no-gpu** |  | **with-gpu** |  |
|---|---|---|---|---|
|  | original | grouping | original | grouping |
| **Encoder** | 34964 (s) |  | 25942 (s) |  |
| **Decoder** | 28430 (s) | 23720 (s) | 8537 (s) | 6906 (s) |

**Table 4.** Computational time comparing grouping scheme and no-grouping scheme in both encoder and decoder on dataset Sideboard

|  | **no-gpu** |  | **with-gpu** |  |
|---|---|---|---|---|
|  | original | grouping | original | grouping |
| **Encoder** | 85413 (s) |  | 57462 (s) |  |
| **Decoder** | 53540 (s) | 45590 (s) | 17120 (s) | 14459 (s) |

**Table 5.** Computational time comparing grouping scheme and no-grouping scheme in both encoder and decoder on dataset Danger_de_Mort

*3) Rate Distortion performance*

To evaluate the impact of transmitting additional information of the residual signals, which is necessary for the reconstruction of the original signals at the Decoder side from its prediction sent from Encoder, this section assesses rate-distortion performance of the proposed super-rays grouping scheme.

Fig. 11 compares rate-distortion performance of the proposal against the top-left view projection graph-based Light Field coding [17] (original) without super-rays grouping, center-view / multiple-view projection graph-based Light Field coding [39] with super-rays grouping, and two state-of-the-art coders: HEVC - Serpentine scanning, JPEG Pleno (4D Transform mode).

The purple line indicates performance of the super-rays grouping proposal for top-left view projection, whereas the blue line represents the original scheme [17] without grouping. It can be seen that datasets having higher number of grouped super-rays like Greek and Fountain_Vincent_2 suffered a slightly higher increase of bitrates (BPP) even though its quality (PSNR) was preserved, and thus their purple lines are lower than blue lines, compared to Sideboard's and Danger_de_Mort's. However, their rate-distortion still tended to perform well at low bitrates, as opposed to HEVC and JPEG Pleno.

The red line indicates how well the super-rays grouping scheme performed when applied on center-view / multiple-view projection (denoted as CVP and MVP respectively) for graph-based Light Field coding [39]. The figures reveal that this approach could outperform all other methods on Fountain_Vincent_2 and Danger_de_Mort, especially at low and high bitrates. However, for Greek and Sideboard, the proposals could only considerably outperform the original scheme [17] at all bitrates, whereas HEVC and JPEG Pleno surpassed their rate-distortion. This can be explained by the fact that synthetic Light Field scenes are free of image imperfections like noises, hence not affecting performance of HEVC and JPEG Pleno, leading to better performance than on real Light Field. Nevertheless, the super-rays grouping scheme proposed for both graph-based approaches remained having the tendency to perform better at low bitrates than HEVC and JPEG Pleno.

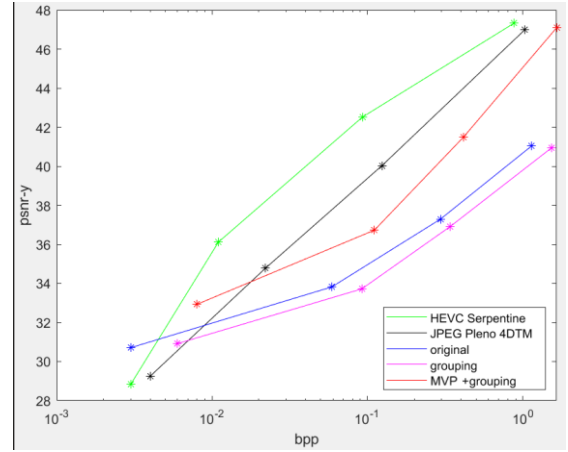

*a) Greek*

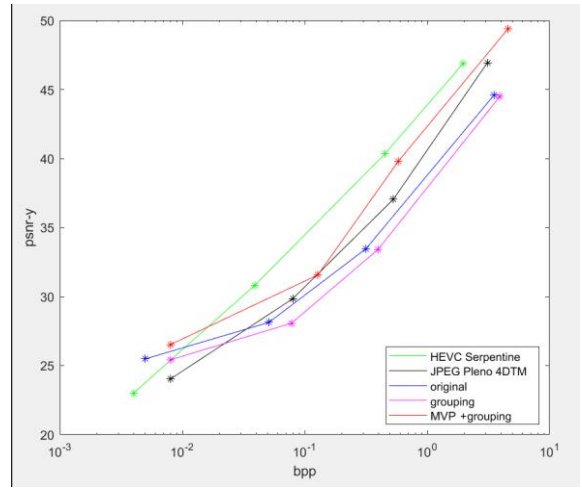

*b) Sideboard*

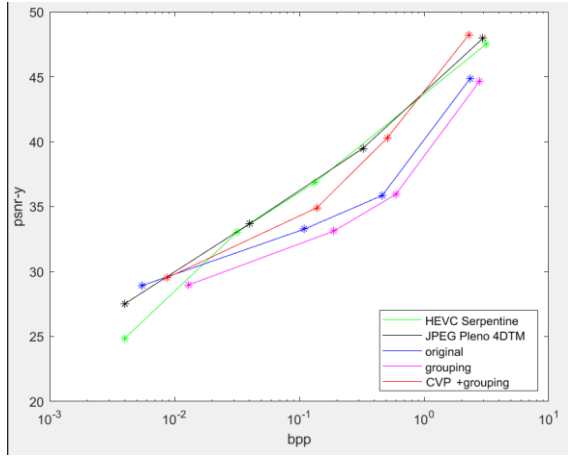

*c) Fountain_Vincent2*

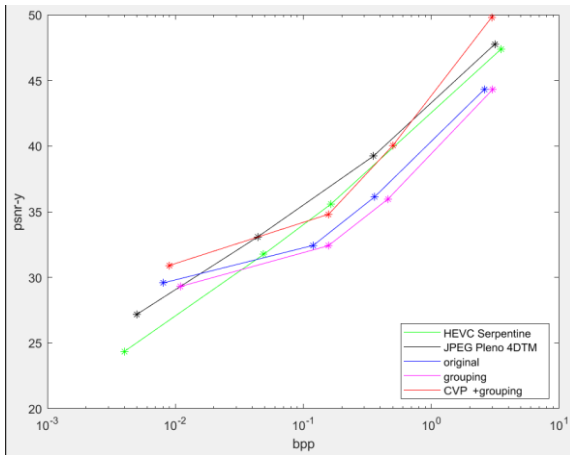

*d) Danger_de_Mort*

**Figure 11.** Rate distortion performance of the proposed super-rays grouping scheme, super-ray grouping with suitable projection schemes [39], original no-grouping scheme and HEVC, JPEG-Pleno standards on four datasets: (a) Greek (b) Fountain_Vincent_2 (c) Sideboard (d) Danger_de_Mort

## V. Discussion

Overall, the proposed super-rays grouping scheme can reduce a considerable amount of decoding time for both simple and complex Light Field scenes, despite raising a slight increase in bitrates, as shown in the rate distortion performance of the proposed coding architecture. Interestingly, super-rays which are not only local but also far from each other, are grouped together based on the similarity of their graph frequency coefficients. As a result, the color signals residing on these super-rays were also found to be similar, leading to small residual data. The extra information per grouped super-ray to be encoded and sent is the residual between its original and predicted signal, which may be further compressed in future work by trying different representations of residual signals or coding using other standards. Additionally, applying the grouping scheme to center-view / multiple view projection of super-rays seems to maintain the potential of graph-based Light Field coding in having superior rate-distortion performance over state-of-the-art methods at low bitrates.

Graph coarsening plays a role in guaranteeing the same dimension for super-rays, which is essential for some local graphs to share similar graph properties like basis functions and have similar reconstructed signals after I-GFT on the same coefficients, as shown in Eq. (3) and (4). The benefit of graph coarsening is also clear for coding low bitrates. At low bitrate coding, most super-rays are coarsened instead of being partitioned. Graph coarsening retains total variations of signal residing on the reduced graph, while reducing substantially the number of coefficients to be coded, resulting in good Rate Distortion at low quality coding. Additionally, real Light Field scenes might also suffer from image noises, limiting the performance of conventional standards like HEVC and JPEG Pleno 4DTM considerably, while not affecting graph coarsening due to utilizing approximation of low rank model. Further Rate Distortion evaluations on other real Light Field scenes with more existing coding methods should be included in our future work to verify this advantage of graph-based approach.

However, at high target quality coding, the number of partitioned super-rays is significantly higher than the coarsened. Hence, the grouping scheme applied only on coarsened super-rays could not perform efficiently, as indicated in overall grouping ratios. Intuitively, one can also group similar partitioned super-rays if their local graphs share approximately the same size and spectral properties. However, since partitioned local graphs are different in dimensions, it is likely to obtain a high number of final groups with different super-ray sizes. Hence, the dimension thresholds should be selected in a way that the number of resulting final groups are insignificant in comparison with the original number of super-rays, but also maintain small residual data. This approach can be further discussed in future work.

Although GPU acceleration for eigen-decomposition has been utilized with cuSOLVER algorithm in Scikit-cuda library and achieved significant time reduction for both Encoder and Decoder, the overall processing time for graph-based Light Field coding remains relatively high, in comparison with existing standard coders. For instance, HEVC and JPEG Pleno decodes the Greek dataset within a few minutes, whereas graph-based approach with our proposed super-rays grouping solution takes approximately 1.8 hours. With the current setup environment, the GPU implementation reduced diagonalization time of a 5000x5000 matrix (one super-ray) from 15 seconds to only 2 seconds on average. However, recent fast Graph Fourier Transform (FGFT) technique has been reported to reduce eigen-decomposition and transform time by a factor of up to 27 based on approximation of basis functions [48, 49], and thus it is possible to implement FGFT for the graph-based Light Field coding to achieve competitive processing time with other standards in future research.

## VI. Conclusion

In this paper, a coarsened super-ray grouping scheme has been proposed along with a novel coding architecture handling extra residual data of signals arising for each super-ray group, to reduce the number of times necessary for eigen-decompositions, leading to computational time reduction for graph-based Light Field compression. This high complexity operation is carried out only once for each main super-ray per group, instead of for every super-ray in that group. The super-rays are grouped based on the similarity of their GFT coefficients at the encoder side, in such a way that when I-GFT using the basis functions of the main super-ray for each remaining one in a group, the predicted signal for each remaining super-ray is as close to the original signal as possible. This proposal has shown a substantial reduce in decoding time for both simple and complex Light Field scenes, when compared with the original non-grouping approach, and remained having competitive Rate Distortion performance to standard coders. Nevertheless, standard coders remain outperforming graph-based approach in computational time, and thus further grouping of partitioned super-rays or the use of Fast Graph Fourier Transform might be taken into consideration in the future research, to achieve better overall grouping ratio at high target quality coding, and feasibility of real-time performance.

## CONFLICT OF INTEREST

The authors declare no conflict of interest.

## AUTHOR CONTRIBUTIONS

Nguyen Gia Bach conceptualized, designed super-rays grouping scheme and Light Field coding architecture, and executed several parts of experiments, then contributed to writing the original draft manuscript. Chanh Minh Tran and Nguyen Duc Tho reviewed related studies, supported running several parts of experiments, and contributed to writing original draft manuscripts. Phan Xuan Tan and Eiji Kamioka supervised, then reviewed and edited the manuscript. All authors have read and agreed to the published version of the manuscript.